\documentstyle[11pt,epsfig]{article}

\def\ct#1{{\cal #1}}
\def\dfrac#1#2{{\displaystyle {#1 \over #2}}}
\def\spose#1{\hbox to 0pt{#1\hss}}
\def\ltapprox{\mathrel{\spose{\lower 3pt\hbox{$\mathchar"218$}}
 \raise 2.0pt\hbox{$\mathchar"13C$}}}
\def\gtapprox{\mathrel{\spose{\lower 3pt\hbox{$\mathchar"218$}}
 \raise 2.0pt\hbox{$\mathchar"13E$}}}
\def\inapprox{\mathrel{\spose{\lower 3pt\hbox{$\mathchar"218$}}
 \raise 2.0pt\hbox{$\mathchar"232$}}}

\newcommand{\Pj}{\mbox{I}\!\!\mbox{P}}
\newcommand{\id}{\mbox{1$\!\!$I}}

\newcommand{\<}{\langle}
\renewcommand{\>}{\rangle}

\newcommand{\SU}{\mbox{SU}} 
\newcommand{\Tr}{\mbox{Tr}\;} 
\newcommand{\RI}{\mbox{\scriptsize RI}}

\newcommand{\sub}{\mbox{\scriptsize sub}}
\newcommand{\latt}{\mbox{\scriptsize latt}}
\newcommand{\phys}{\mbox{\scriptsize phys}}
\newcommand{\tree}{\mbox{\scriptsize tree}}
\newcommand{\QCD}{\mbox{\scriptsize QCD}}
\newcommand{\be}{\begin{equation}}
\newcommand{\ee}{\end{equation}}
\newcommand{\bea}{\begin{eqnarray}}
\newcommand{\eea}{\end{eqnarray}}

\begin{document}
\hskip 1cm\\
\vskip 3cm
\leftline{\bf WEAK MATRIX ELEMENTS ON THE LATTICE:}
\vskip 0.2cm
\leftline{\bf RECENT DEVELOPMENTS IN $K$-PHYSICS}
\vskip 1.5cm
\hskip .7in \mbox{\bf M. Talevi}
\vskip 0.3cm
\hskip .7in \mbox{Dip. di Fisica, Univ. di Roma ``La Sapienza'' and 
INFN, Sezione di Roma}
\vskip 0.1cm
\hskip .7in \mbox{P.le A. Moro 2, I-00185 Roma, Italy.}
\vskip 1.5cm
 
\subsubsection*{INTRODUCTION}
\label{sec:introduction}
 
In this talk we present some recent developments in the calculation of weak 
matrix elements on the lattice.  Lattice QCD is one of the few 
systematically improvable methods for computing them from first principles, and
has proven a powerful and appealing approach.
In spite of the successes, progress has been slow due to the presence of
systematic effects, such as discretization and non-perturbative renormalization
effects.  In the following, we concentrate on the 
applications in $K$-physics of a recently introduced method for 
non-perturbative (NP) renormalization \cite{NPM}.

Renormalization of lattice operators is a crucial ingredient in the calculation
of physical weak matrix elements on the lattice.
A physical amplitude $A_{\alpha \rightarrow \beta}$ of a weak transition
$\alpha \rightarrow \beta$ is calculated via the Operator Product Expansion 
(OPE) by
\begin{equation}
A_{\alpha \rightarrow \beta} = C_W(\mu/M_W) \< \alpha | \hat O(\mu) |\beta \>
\label{eq:ope}
\end{equation}
where $C_W$ is the Wilson coefficient of the OPE, $M_W$ is the mass of the
$W$ boson, 
$\mu$ is the renormalization scale and $\<\alpha|\hat O(\mu)|\beta \>$ is 
the matrix element of the renormalized operator (at the scale $\mu$) relevant 
to the physical process.
The Wilson coefficient $C_W(\mu/M_W)$ contains the short-distance information 
and can be calculated in Perturbation Theory (PT) in the continuum at the 
renormalization scale $\mu$. 
The matrix element contains the long-distance dynamics and thus must 
be calculated non-perturbatively on the lattice.
Renormalization relates the regularized lattice matrix elements 
to its continuum counterpart.  

On the lattice, chiral symmetry is explicitly broken with Wilson-like fermions.
The possibility of recovering the chiral symmetry in the continuum limit 
was shown in \cite{Bochicchio}.  The general prescription is to subtract 
from the bare operator $O(a)$ all the operators of dimension less or equal
than $O(a)$, which have the same quantum numbers left unbroken by the 
regularization:
\begin{equation}
\< \alpha | \hat O(\mu) | \beta \> = \lim_{a\to 0}
\< \alpha | Z_O(\mu a)[O(a)+\sum_i Z_i O_i(a)] |\beta \>,
\end{equation}
If the subtracted operators $O_i$ have lower dimension than $O$, 
the mixing constants are power-divergent in the 
cutoff, $Z_i\sim 1/a^d$ with $d>0$.   These divergent factors can pick up
exponentially small contributions in the strong coupling, yielding a finite
contribution as $a\to 0$, i.e. 
\begin{equation}
\frac{1}{a} e^{-1/\alpha_s(a)}\sim \Lambda_{\QCD}.
\end{equation} 
These divergences must be subtracted in a completely non-perturbative way.
Recently, non-perturbatively renormalization has witnessed a great
progress \cite{NPM,ALPHA,JLQCD}.  Here, we want to discuss the applications
of the method of ref. \cite{NPM}, which in the following will be refered to as
the NP method (NPM).

We will concentrate here on two physical processes of phenomenological interest:
\begin{itemize}
\item  $K^0 - \bar K^0$ mixing, in which the bare operator
$O^{\Delta S = 2}={\bar s} \gamma^L_{\mu} d {\bar s} \gamma^L_{\mu}  d$ 
mixes only with operators of the same dimension \cite{DS=2,NP_pm}.  

From the matrix element of the $\Delta S = 2$ operator
the kaon $B$-parameter $B_K$ is obtained, which is a 
quantity of great phenomenological interest,
being related to the $\epsilon$-parameter which measures CP-violation 
in the $K^0$--$\bar K^0$ system.  With the measured value of the top quark
mass, an accurate prediction of $B_K$ enables us to limit
the range of values of the CP-violation phase $\delta$.

\item $K \rightarrow \pi \pi$ decay, in which 
the bare operators $O^{\pm}_{LL}$ also mix with operators of lower 
dimensionality \cite{DI=1/2}.

These decays are relevant to the study of $\Delta I=1/2$ rule, 
the enhancement of the $\Delta I=1/2$ with respect to the $\Delta I=3/2$ rate,
a long-standing question which lattice QCD should be able to understand 
in a quantitative fashion.  Yet, this has proven a formidable task due to the 
power-divergent mixing.  The NPM allows us to make the subtraction without
loosing predictive power, as one would do imposing renormalization conditions
directly on hadronic states \cite{Maiani}.
\end{itemize}

\subsubsection*{NON-PERTURBATIVE METHOD}

In the NPM, the renormalization conditions are applied directly to the 
Green functions of quarks and gluons, in a fixed gauge, with given off-shell
external states of large virtualities \cite{NPM}.  
The method mimicks what is usually
done in the perturbative calculation, but the Green functions are evaluated
in a non-perturbative fashion from Monte Carlo simulations.

To give the flavour of the method, let us consider the simplified case of a 
multiplicatively renormalizable operator, 
e.g. a two-quark operator $O=\bar q \Gamma q$.
Given the bare lattice operator $O^{\latt}(a)$, the renormalization condition
we impose is \cite{NPM} 
\begin{equation}
Z^{\latt}_{\RI}(\mu a)\<p|O^{\latt}(a)|p\>|_{p^2=\mu^2}=
\<p|O^{\latt}(a)|p\>|_{p^2=\mu^2}^{\tree},
\label{eq:Z_RI(mu)}
\end{equation}
where $\<p|\cdots|p\>$ denotes the matrix element of external quarks of momenta
$p$ which can be calculated to all orders in the QCD coupling via Monte Carlo
simulations.  
The renormalized operator obtained with the NPM is then
\begin{equation}
\hat O_{\RI}(\mu)=Z^{\latt}_{\RI}(\mu a)O^{\latt}(a),
\label{eq:O_RI(mu)}
\end{equation}
which depends on the external states and the gauge, but not on method used to 
regulate the ultra-violet divergences.  
To stress this point, we call the NP renormalization scheme 
Regularization Independent (RI) \cite{eps/eps'}.  The physical operator 
\begin{equation}
O^{\phys}(M_W)=C_{\RI}(M_W/\mu)\hat O_{\RI}(\mu)
\label{eq:O^phys}
\end{equation}
is independent of external momenta and gauge 
(up to higher orders in continuum PT and lattice systematic effects) if the
Wilson coefficient function $C_{\RI}(M_W/\mu)$ in the RI scheme is calculated 
with the same external momenta and gauge of $\hat O_{\RI}(\mu)$. 
The advantage of the RI scheme is that it completely avoids the use of 
lattice PT, which is expected to have a worse convergence than the continuum
expansion \cite{Lepage}.  The coefficient function 
$C_{\RI}(M_W/\mu)$ are instead calculated in continuum PT, 
which cannot be avoided since the Wilson OPE is defined perturbatively.  

The NPM is valid for any composite operator, as long 
as we can can find a window in the range of $\mu$ such that 
$\Lambda_{\QCD}\ll\mu\ll O(1/a)$, in order to keep under control both the
higher-order effects in the (continuum) perturbative calculation of $
C_{\RI}$ and discretization errors \cite{NPM}.  
We stress that this requirement is common to all NP methods on the lattice.

\subsubsection*{$\Delta S=2$}

We consider the renormalization of the four-fermion operator
\begin{equation}
O^{\Delta S=2}=(\bar s \gamma_{\mu}^L d )(\bar s \gamma_{\mu}^L d)
\, , \ \gamma_{\mu}^L=\frac{1}{2}\gamma_{\mu} (1-\gamma_5)\, , 
\label{eq:O_DS=2} 
\end{equation}
which appears in  the weak effective Hamiltonian relevant
for $K^0$--$\bar K^0$ mixing.

The Wilson term in the quark action induces the
mixing of the operator (\ref{eq:O_DS=2}) with dimension-six operators of 
different chirality, determined by CPS symmetry of the action
(the S stands for the $s\leftrightarrow d$ flavour exchange symmetry) 
\cite{BDHS,DS=2,NP_pm}
\begin{equation}
\begin{array}{lcl}
O_1&\equiv&-\dfrac{1}{16N_c} [ O_{SS} - O_{PP}  ] \\
O_2&\equiv&-\dfrac{(N_c^2+ N_c-1)}{32N_c} [ O_{VV} - O_{AA} ] \\
O_3&\equiv& \dfrac{( N_c-1)}{16N_c} [ O_{SS} + O_{PP} + O_{TT} ] \\
O_4&\equiv&\dfrac{(N_c-1)}{16N_c} [ O_{SS} + O_{PP} -\frac{1}{3} O_{TT} ]
\end{array}
\label{eq:oo}
\end{equation}
where $N_c$ denotes the number of colours and
$O_{\Gamma\Gamma}=(\bar s \Gamma d )(\bar s \Gamma d)$, with $\Gamma$ one of the
Dirac matrices.  Note that $O_4$ is not present in 1-loop PT \cite{PT_pm}, 
so the mixing is at most of $\ct{O}(g_0^4(a))$, 
where $g_0(a)$ is the bare lattice QCD coupling.
We define the renormalized operator as
\begin{equation}
\hat O^{\Delta S=2}_{\RI}(\mu)
=Z^{\Delta S=2}(\mu a)O^{\Delta S=2}_{\sub}(\mu a)  
\equiv Z^{\Delta S=2}(O^{\Delta S=2}+\sum_{i=1}^{4}Z_i O_i) 
\label{eq:O_DS=2(mu)}
\end{equation}
Since there are no $\Delta S=2$ operators of dimension lower than six,
the mixing constants $Z_i$ are finite, while the logarithmically divergent 
$Z^{\Delta S=2}$ renormalizes multiplicatively the subtracted
operator $O^{\Delta S=2}_{\sub}$.

\begin{figure}   % produce figure here
\vspace{0.1cm}
%\begin{center}\setlength{\unitlength}{1mm}
%\begin{picture}(160,100)
%\put(10,-35)
%{\special{PSfile=zmix_k1432.ps}}
%\end{picture}
%\end{center}
\centerline{\epsfig{figure=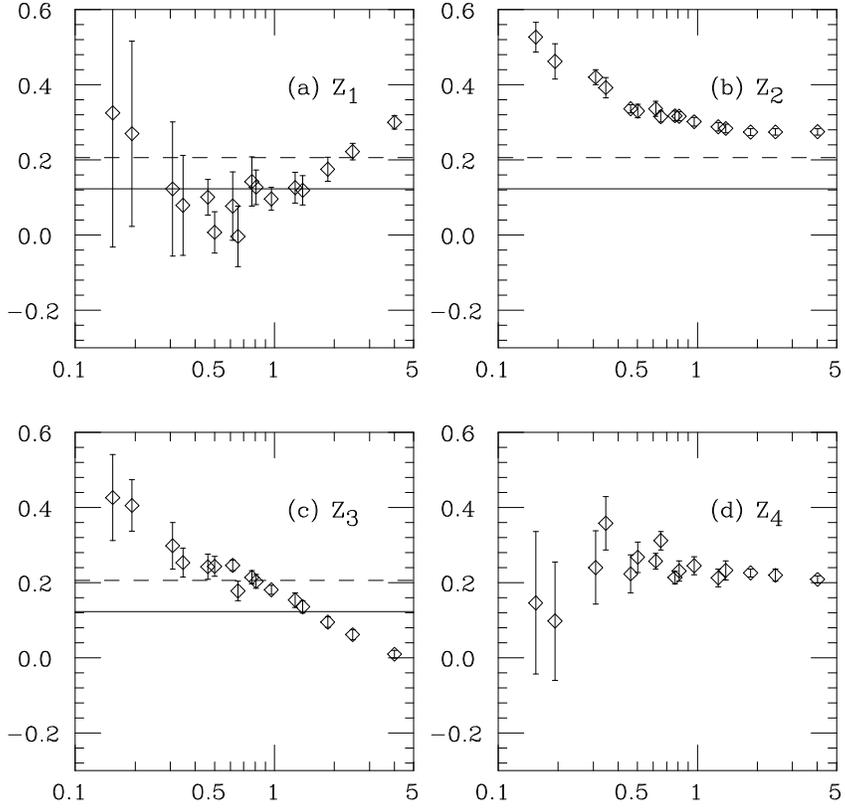,height=15cm,angle=90}}
\caption{Mixing constants $Z_i (i = 1,\dots,4)$ at $\kappa=0.1432$, 
for several renormalization scales $\mu^2a^2$.  
The solid line is the result from ``standard'' PT, while the dashed line comes
from ``boosted'' PT with $\alpha_V\simeq\ 1.68\ \alpha_s^{\latt}$.}
\label{fig:Z_i}
\end{figure}

We determine the mixing constants $Z_i$ with a projection method on the 
four-point amputated Green functions  \cite{DS=2,NP_pm}.  Denoting by 
$O_0$ the operator $O^{\Delta S=2}$, we define a set of mutually orthogonal
projectors $\Pj_i\ i=0,\ldots,4$ on the amputated tree-level four-quark
Green functions $\Lambda_i^{(0)}$ of the operators $O_i$, by the relation 
$\Tr\Pj_i\Lambda_j^{(0)}=\delta_{ij}$, where the trace over colour and spin
is understood.  The mixing constants $Z_i$ are fixed by the condition that the 
subtracted operator $O^{\Delta S=2}_{\sub}$ be proportional to the bare free 
operator, i.e.
\begin{equation}
\Tr\Pj_k \Lambda^{\Delta S=2}_{\sub}(pa)=0,\ k=1,\ldots,4,
\label{eq:mixing_condition}
\end{equation}
where $\Lambda^{\Delta S=2}_{\sub}(pa)$ is the amputated Green function of 
$O^{\Delta S=2}_{\sub}$, which is calculated at equal
external momenta $p$ and in the Landau gauge in a completely NP fashion from 
numerical simulations.  Eq. (\ref{eq:mixing_condition}) yields
a linear non-homogeneous system, from which we determine the mixing 
constants $Z_i$.  Once we have determined the mixing constants, 
the overall renormalization constant $Z^{\Delta S=2}$ is determined by 
\cite{DS=2,NP_pm}
\begin{equation}
Z^{\Delta S=2}(\mu a)Z_q^{-2}(\mu a)
\Gamma^{\Delta S=2}_{\sub}(pa)|_{p^2=\mu^2}=1,
\label{eq:Z_+}
\end{equation}
where 
$\Gamma^{\Delta S=2}_{\sub}(pa)= \Tr \Pj_0 \Lambda^{\Delta S=2}_{\sub}(pa)$,
and $Z_q$ is the light-quark renormalization constant, determined from 
the conserved vector current in a NP way \cite{NPM}.

The quark Green functions have been calculated using Monte Carlo simulations.
We have used an ensemble of 100 configurations, on a $16^3\times 32$ lattice, 
at $\beta=6.0$, and with three values of the hopping parameter (which is related
to the quark mass) 
$\kappa=0.1425,0.1432,0.1440$  for the quark propagator, in the lattice Landau
gauge.  To reduce lattice artefacts, we have used an $\ct{O}(a)$ 
improved lattice quark action \cite{Heatlie}.
We have performed the calculation for a wide range of scales $\mu^2a^2$.
In fig.\ \ref{fig:Z_i}, we show the NP mixing constants at $\kappa=0.1432$ 
as a function of $\mu^2a^2$,  and compare 
them with the PT result in the same gauge and external momenta \cite{NP_pm}.  
We also report the PT result using a ``boosted'' coupling 
$\alpha_V\simeq\ 1.68\ \alpha_s^{\latt}$ \cite{Lepage}. 
We have checked that the dependence of the $Z$'s on $\kappa$ is very mild.
As expected the higher-order contributions differentiate the 
mixing constants with respect to the PT value, which at one-loop is the same
for all the constants \cite{PT_pm}. 
We note that $Z_2$ and $Z_4$ are very well defined and almost scale
independent in a large ``window'' of $\mu^2a^2$, whereas $Z_1$ and $Z_3$
are more scale-dependent.  Moreover, $Z_4$ which is absent in 1-loop
PT is not neglegible.

The effects of the NP corrections can be most clearly seen in the study
of the chiral behaviour of the matrix element
$\<\bar K^0|\hat O^{\Delta S=2} | K^0\>_{\latt}$.  The bare matrix elements 
have been computed from an ensemble of 460 configurations with the same
$\ct{O}(a)$ improved action, on a $18^3\times 64$
lattice, at $\beta=6.0$ \cite{B_K}.  Parametrizing the lattice 
matrix element as  
\begin{equation}
\<\bar K^0(p)|\hat O^{\Delta S=2}_{\RI} | K^0(q)\>
=\alpha+\beta m_K^2+\gamma(p\cdot q)+...
\label{eq:parameters}
\end{equation}
one finds the values of $\alpha,\beta$ and 
$\gamma$ at several renormalization scales $\mu^2 a^2$, cf.
tab. \ref{tab:param}.  The $B_K$ parameter in the RI scheme is given by
$B_K^{\RI}=\gamma(\mu)/Z_A^2$, where $Z_A=1.06(3)$ is the axial current 
renormalization constant.  $\alpha$ is a lattice artefact and should 
vanish in the continuum limit.

\begin{table}
\centering
\begin{tabular}{rrrr}
\hline
$\mu^2 a^2$ & $\alpha$ & $\beta$ & $\gamma$ \\ \hline 
0.31 & $ 0.030(18)$ &  0.27(21) & 0.90(15) \\ 
0.62 & $-0.027(16)$ &  0.36(18) & 0.75(13) \\ 
0.96 & $-0.012(14)$ &  0.24(17) & 0.69(12) \\ 
1.27 & $ 0.005(13)$ &  0.14(16) & 0.68(12) \\ 
1.39 & $-0.009(13)$ &  0.24(16) & 0.67(12) \\ 
1.85 & $-0.003(13)$ &  0.18(16) & 0.66(11) \\ 
2.46 & $-0.001(12)$ &  0.24(15) & 0.65(11) \\ 
4.01 & $-0.002(12)$ &  0.44(15) & 0.67(11) \\ 
 BPT & $-0.052(12)$ &  0.16(15) & 0.62(11) \\ \hline
\end{tabular}
\caption{Values of $\alpha,\beta$ and $\gamma$ for several
renormalization scales $\mu^2 a^2$. In the last row we present results
obtained with ``boosted'' PT renormalization constants.}
\label{tab:param}
\end{table}
As can be seen, in the window of $\mu^2 a^2$ values in which the $Z$'s were 
found to be reliable, $\alpha$ is compatible with zero within one standard 
deviation, whereas $\beta$ is compatible with zero within 1.5 standard 
deviations. On the other hand, in the same window of $\mu^2 a^2$, $\gamma$ 
stabilizes to a non-zero value.   The fluctuations of $Z_1$ and $Z_3$ do not
influence the stability of the results since the relative matrix elements
$\<\bar K^0(p)|O_{1,3}| K^0(q)\>$ are quite small compared to
$\<\bar K^0(p)| O_2 | K^0(q)\>$,  whose mixing constant $Z_2$ is extremely well
determined.
The vanishing of $\alpha$ allows us to conclude that the use of the NP $Z$'s 
improves the chiral behaviour for a large range of values of $\mu^2a^2$ 
\cite{NP_pm}. 

One we are confident that the correct chiral behaviour is recovered, we can
reliably calculate the physical value of $B_K$, obtained by calculating the
Wilson coefficient in the same RI scheme \cite{B_K,NP_pm}. 

\subsubsection*{$\Delta I=1/2$}

In the continuum, with an active charm quark and the GIM mechanism at work,
the operator basis is given by \cite{Maiani}
\begin{equation}
O^{\pm}_{LL}= \frac{1}{2}[(\bar s d)_L(\bar u u)_L
                      \pm (\bar s u)_L(\bar u d)_L]   -(u\to c).
\label{eq:OpmGIM}
\end{equation}
Again, the renormalization strategy is complicated by chiral symmetry breaking.
In fact, the Wilson term induces the mixing of $O^{\pm}_{LL}$ with 
lower-dimensional operators, with power-divergent coefficients, 
which need to be subtracted non-perturbatively.  
In order to renormalize the operators (\ref{eq:OpmGIM}) on the lattice 
we find it convenient to separate the $\{8,1\}$ and the $\{27,1\}$ components
under the $\SU(3)_L\otimes\SU(3)_R$ chiral group.
In the following, we shall concentrate only on the octet component of 
$O^{\pm}_{LL}$, which we will denote with $O^{\pm}_0$ 
\cite{DI=1/2}\footnote{The lattice penguin operators, being 
proportional to $(m_c^2-m_u^2)a^2\ll 1$, will be neglected in the following.}. 

There are different ways of calculating the $K\to\pi\pi$ matrix elements,
which correspond to different renormalization structures.
We considering here a general structure of the form
\begin{equation}
\widehat O^{\pm}=Z^{\pm}\left[O^{\pm}_0+\sum_{i=1}^4 Z^{\pm}_iO^{\pm}_i
+Z^{\pm}_5O_5+Z^{\pm}_3O_3\right],
\label{eq:hatOpm}
\end{equation}
refering to \cite{DI=1/2} for a more detailed analysis.
In eq.~(\ref{eq:hatOpm}) $O^{\pm}_0$ are the bare operators,
$O^{\pm}_i,\ i=1,\ldots,4$ are dimension-six operators of wrong 
chirality (similarly to the $\Delta S=2$ case), 
$O_5$ is a dimension-five of the form 
$\bar s\Sigma_{\mu\nu}F_{\mu\nu}d$ ($\Sigma_{\mu\nu}=\sigma_{\mu\nu}$ or 
$\tilde\sigma_{\mu\nu}$) and $O_3$ is a dimension-three operator of the form
$\bar s\Gamma d$ ($\Gamma=\id$ or $\gamma_5$). 

According to the NPM, the mixing $Z$'s are determined by finding a set of 
projectors on the tree-level amputated Green functions (GF), with off-shell
quark and gluon external states,
the choice of which depends on the nature of the operators at hand.
For the $\Delta I=1/2$ operators we choose the following set of external 
states: $q\bar q$, $q\bar qg$, $q\bar qq\bar q$, with the momenta given 
below in eq.~(\ref{eq:6-Z-system}).  
For each choice of external states, i.e. for each different set of GF,
we need different type of projectors.  Let us denote with $\Pj_3$ 
the projector on the $q\bar q$ GF of the operator $O_3$, 
with $\Pj_5$ the projector on the $q\bar qg$ GF of the 
operator $O_5$, and with $\Pj_j,\ j=1,\ldots,4$ the set of mutually
orthogonal projectors on the operators $O_i,\ i=1,\ldots,4$.  We refer the
reader to refs.\ \cite{DI=1/2,NP_pm} for the explicit expressions of the
projectors.  Applying the projectors to the corresponding NP GF
of the renormalized operators $\widehat O^{\pm}$,
with an appropriate choice of the external states, 
we require that the renormalized operators be proportional to the bare 
operators, $\widehat O^{\pm}(\mu)\propto O^{\pm}_0(a)$ (up to terms of 
${\cal O}(a)$), i.e.\ we impose the following renormalization conditions
(trace over colour and spin is understood in the projection operation):
\begin{equation}
\begin{array}{l}
\Pj_3\<q(p)|\widehat O^{\pm}|\bar q(p)\>=0  \\
\Pj_5\<q(p-k)g(k)|\widehat O^{\pm}|\bar q(p)\>=0  \\
\Pj^{\pm}_j\<q(p)\bar q(p)|\widehat O^{\pm}|q(p)\bar q(p)\>=0,\ j=1,\ldots,4  
\end{array}
\label{eq:6-Z-system}
\end{equation}
where $p$ and $k$ denote the momentum of the external quark and gluon legs.
The system of equations (\ref{eq:6-Z-system}) completely determines in a NP 
way the renormalization constants, as we have six conditions 
(non-homogeneous due to the matrix elements of $O^{\pm}_0$, 
cf.~eq.~(\ref{eq:hatOpm}))
in six unknown mixing constants, $Z^{\pm}_i,\ i=1,\ldots,4,Z^{\pm}_5,Z^{\pm}_3$.

Unfortunately, since solving eq.~(\ref{eq:6-Z-system}) involves
delicate cancellations between large contributions, it may very likely result
in a very noisy determination, even with large statistics.  
An equivalent strategy we can adopt is:\\
1.\ We introduce an intermediate subtraction for the dimension-five 
and -six operators
\begin{equation}
\begin{array}{l}
\bar O^{\pm}_i=O^{\pm}_i+C^{(\pm,i)}_3 O_3 ,\ i=0,\ldots,4, \\
\bar O_5=O_5+C^{(5)}_3 O_3,
\end{array}
\end{equation}
and determine the power-divergent mixing constants $C^{(\pm,i)}_3$ and 
$C^{(5)}_3$ by imposing 
\begin{equation}
\begin{array}{l}
\Pj_3\<q(p)|\bar O^{\pm}_i|\bar q(p)\> =0,\ i=0,\ldots,4, \\
\Pj_3\<q(p)|\bar O_5|\bar q(p)\> =0.
\end{array}
\end{equation}
2.\ The finite mixing constants $Z^{\pm}_i$ and $Z^{\pm}_5$,
which in principle can be calculated in perturbation theory (PT),
are then determined from the system 
\begin{equation}
\begin{array}{l}
\Pj_5\<q(p-k)g(k)|\sum_iZ^{\pm}_i\bar O^{\pm}_i
                 +Z^{\pm}_5\bar O_5|\bar q(p)\> 
=\Pj_5\<q(p-k)g(k)|\bar O^{\pm}_0|\bar q(p)\> \\
\Pj_j\<q(p)\bar q(p)|\sum_iZ^{\pm}_i\bar O^{\pm}_i
                    +Z^{\pm}_5\bar O_5|q(p)\bar q(p)\> 
=\Pj_j\<q(p)\bar q(p)|\bar O^{\pm}_0|q(p)\bar q(p)\>,\ j=1,\ldots,4.
\end{array}
\end{equation}
The numerical implementation of this program is in progress \cite{DI=1/2}.

\subsubsection*{CONCLUSIONS}

Over the last few years there has been considerable progress in the 
application of lattice QCD to calculate weak matrix elements.
Much control has been gained over the two major systematic errors 
one had to face, discretization effects
and higher orders effects in the lattice perturbative expansion.
Discretization effects have been greatly reduced
by the ``improvement'' program \cite{Heatlie,Luscher}, while the results 
obtained with the non-perturbative renormalization program make us confident 
that we can realiably calculate weak matrix elements in the low-energy regime.

\subsubsection*{ACKNOWLEDGEMENTS}

The results presented here have been obtained in a most enjoyable and fruitful
collaboration with A. Donini, V. Gim\`enez, G. Martinelli, G.C. Rossi, 
C.T. Sachrajda, M. Testa and A. Vladikas, to whom I am much indebted for all 
that I've learned.

\end{document}